\documentclass[10pt, conference]{IEEEtran}

\usepackage{listings}

\usepackage{environ}
\usepackage{array, color}
\usepackage{verbatim}

\usepackage{enumitem}
\usepackage{soul}
\usepackage{color}
\usepackage{url}
\usepackage{xspace}
\usepackage{booktabs}
\usepackage{tabularx}
\usepackage{multirow}
\usepackage[normalem]{ulem}
\usepackage{anyfontsize}
\usepackage[table]{xcolor}
\useunder{\uline}{\ul}{}
\usepackage{balance}
\usepackage{graphicx}
\usepackage[ruled,vlined,procnumbered]{algorithm2e}
\usepackage{listings} 
\usepackage{microtype}
\usepackage[tight,footnotesize]{subfigure}
\usepackage{subfigure}
\usepackage{pslatex}
\usepackage{seqsplit}
\usepackage{amssymb}
\usepackage{amsmath}
\usepackage{algorithmic}
\usepackage{listings}
\newcommand{\algsmall}{
	\algsetup{linenosize=\small}
	\small}

\usepackage{tikz}
\newcommand*\circled[1]{\tikz[baseline=(char.base)]{\textbf{
			\node[shape=circle,draw,inner sep=0.6pt] (char) {#1};}}}

\usepackage{hyperref}

\newcommand{\ARCADE}{\emph{ARCADE}\xspace}
\newcommand{\arcade}{\emph{ARCADE}\xspace}
\newcommand{\acdc}{\emph{ACDC}\xspace}
\newcommand{\ACDC}{\emph{ACDC}\xspace}
\newcommand{\ARC}{\emph{ARC}\xspace}
\newcommand{\arc}{\emph{ARC}\xspace}

\newcommand{\ourappr}{\emph{UnArch}\xspace}

\definecolor{brown}{cmyk}{0,0.81,1,0.60}
\definecolor{magenta}{rgb}{0.4,0.7,0}
\definecolor{gray}{rgb}{0.5,0.5,0.5}
\definecolor{red}{rgb}{1,0,0}
\definecolor{yellow}{rgb}{.85,0.75,0}
\definecolor{purple}{rgb}{0.5,0,0.5}
\definecolor{green}{rgb}{0.0,0.2,0.0}
\definecolor{blue}{rgb}{0,0,1}

\definecolor{lgray}{gray}{0.85}
\definecolor{dgray}{gray}{0.65}

\newcolumntype{L}[1]{>{\raggedright\let\newline\\\arraybackslash\hspace{-4pt}}m{#1}}
\newcolumntype{C}[1]{>{\centering\let\newline\\\arraybackslash\hspace{-5pt}}m{#1}}
\newcolumntype{R}[1]{>{\raggedright\let\newline\\\arraybackslash\hspace{-5pt}}m{#1}}

\makeatletter
\def\@copyrightspace{\relax}
\makeatother

\begin{document}
%
\title{Uncovering Architectural Design Decisions  }

\author{
		\IEEEauthorblockA{Arman Shahbazian, Youn Kyu Lee, Duc Le, and Nenad Medvidovic}
	\\ \vspace{-3mm}
	\setlength{\tabcolsep}{40pt}
	\begin{tabular}{ c  }
		Computer Science Department \\
		University of Southern California\\
		\{armansha, younkyul, ducmle, neno\}@usc.edu \\ \\
	\end{tabular}
}
\maketitle

\begin{abstract}

Over the past three decades, considerable effort has been devoted to the study of software architecture. A major portion of this effort has focused on the originally proposed view of four ``C''s---components, connectors, configurations, and constraints---that are the building blocks of a system's architecture. Despite being simple and appealing, this view has proven to be incomplete and has required further elaboration. To that end, researchers have more recently tried to approach architectures from another important perspective---that of design decisions that yield a system's architecture. These more recent efforts have lacked a precise understanding of several key questions, however: (1) What is an architectural design decision (definition)? (2) How can architectural design decisions be found in existing systems (identification)? (3) What system decisions are and are not architectural (classification)? (4) How are architectural design decisions manifested in the code (reification)? (5) How can important architectural decisions be preserved and/or changed as desired (evolution)? This paper presents a technique targeted at answering these questions by analyzing information that is readily available about software systems. We applied our technique on over 100 different versions of two widely adopted open- source systems, and found that it can accurately uncover the architectural design decisions embodied in the systems.
\end{abstract}

\IEEEpeerreviewmaketitle
\section{Introduction}
Software architecture has become the centerpiece of modern software development, especially with the software development having its focus shifted from lines-of-code to coarser grained architectural elements such as software components and connectors \cite{taylor2009software}. 
Developers are increasingly relying on software architecture to lead them through the process of creating and implementing large and complex systems.

Considerable effort has been devoted to studying software architecture from different perspectives. Kruchten et al. proposed the 4+1 view model to describe software architecture using five concurrent views addressing specific concerns \cite{kruchten19954}. Another research thread has focused on architecture description languages (ADLs) \cite{medvidovic2000classification} such as ACME \cite{garlan2010acme} and $\pi$-ADL \cite{oquendo2004pi} for describing and modeling software systems' architectures. Architectures of the several now widely adopted systems such as mobile systems \cite{bagheri2016software}, and World Wide Web \cite{fielding2002principled} has been extensively studied. Furthermore, researchers have recognized the importance of software architecture in the evolution and maintenance of software systems \cite{garcia2013comparative} which has led to the design of   several architectural recovery techniques to help counteract the challenges brought about by architectural drift and erosion \cite{de2012controlling, garcia2011enhancing, garcia2013comparative, tzerpos2000acdc}.

These studies are typically based on some slant on the originally proposed view of software architecture as four ``C''s---components, connectors, configurations, and constraints---that are the building blocks of a system's architecture. Despite being simple and appealing, this view has proven to be incomplete and has required further elaboration. To that end, researchers have more recently tried to approach architectures from another important perspective---that of design decisions that yield a system's architecture. These more recent efforts have lacked a precise understanding of several key questions, however: \circled{1} What is an architectural design decision (definition)? \circled{2} How can architectural design decisions be found in existing systems (identification)? \circled{3} What system decisions are and are not architectural (classification)? \circled{4} How are architectural design decisions manifested in the code (reification)? \circled{5} How can important architectural decisions be preserved and/or changed as desired (evolution)? This paper presents a technique (\ourappr) targeted at answering these questions by analyzing information that is readily available about software systems.

Our approach is guided and constrained by the following observations. 
	The system already exists.
	We have access to the source code and  issue repositories. We assume that the issue repository contains specific information, i.e., the list of issues and a means to obtain the pertaining code changes such as attached code commits or pull requests.
	Architectural documentation does not exist, is incomplete, or is unreliable.
	Access to the architects of some of the systems may be possible, but it is not likely. Furthermore, the architects' availability is highly constrained. 
	Finally, the architects may not remember or be able to articulate their design decisions.

\ourappr builds on top of the state-of-the-art architecture recovery techniques that recover a descriptive architecture of a system. A major shortcoming of these techniques is that they only depict ``what'' the architecture of a system looks like, and not ``why'' the architecture looks the way it does---which is the symptom of a phenomenon called knowledge vaporization in software systems \cite{jansen2005software}. To overcome this shortcoming, \ourappr taps into the issue and code repositories. Using the information obtained from the code and issue repositories as its inputs, \ourappr outputs a set of decisions that has been made during the system's evolution. 
We applied our technique on over 100 different versions of two widely adopted open- source systems, and found that it can accurately uncover the architectural design decisions embodied in the systems.

The contributions of this paper are defining and classifying different kinds of architectural design decisions that are made during the evolution of a software system; devising a technique to uncover and identify those design decisions in existing software systems; and empirically examining how they are manifested in software systems.

The remainder of this paper is organized as follows. Section \ref{sec:foundation} summarizes the fundamental research thrusts and concepts brought together to enable this work. Section \ref{sec:approach} describes \ourappr in detail. Sections \ref{sec:evaluation} describe our evaluation and key findings. Section \ref{sec:threats} describes the threats to the validity of our approach and its mitigating factors. A discussion of related work (Section \ref{sec:related}) and conclusions (Section \ref{sec:conclusions}) round out the paper.

\section{Foundation}\label{sec:foundation}
\subsection{Architectural Design Decisions}\label{sec:f-d}
For many years research community and industry alike has been focused on the result, the consequences of the design decisions made, trying
to capture them in the ``architecture'' of the system under consideration, often using graphics.
Representations of software architecture were and to a great extent still are centered on views, \cite{clements2002documenting, kruchten19954}, as captured by the  ISO/IEC/IEEE 42010 standard \cite{international2007iso}, or usage of an architecture description language \cite{medvidovic2000classification}. 
However, this approach to documenting software architectures can cause problems such as expensive system evolution, lack of stakeholders communication, and limited re-usability of architectures \cite{shahin2009architectural}. 

Architecture as a set of design decisions was proposed to address these shortcomings. This new paradigm focuses on capturing and documenting rationale, constraints, and alternatives of design decisions.
More specifically Jansen et al. defined architectural design decisions as a description of the set of architectural additions, subtractions and modifications to the software architecture, the rationale, the design rules, and additional requirements that (partially) realize one or more requirements on a given architecture \cite{jansen2005software, bosch2004software}. The key element in their definition is rationale, i.e., the reasons behind an architectural design decision.
Kruchten et al. proposed an ontology that classified architectural decisions into 3 categories: (1) existence decisions (ontocrises), (2) property decisions (diacrises), and (3) executive decisions (pericrises) \cite{kruchten2004ontology}. Among the three categories, existence decisions -- decisions that state some element/artifact will positively show up or disappear in a system -- are the most prevalent and capture the most volatile aspects of a system \cite{kruchten2004ontology, jansen2005software}. Property and executive decisions are enduring guidelines that are mostly driven by the business environment and affect the methodologies, and to a large extent the choices of technologies and tools.

The notion of design decisions used in this paper values the ``rationales'', and ``consequences'' as two equally important constituent part of design decisions. 
However, not all design decisions are created equal. Some design decisions are straight forward with clear singular rationale and consequence while some are crosscutting and intertwined \cite{bosch2004software}, i.e., affect multiple components, connectors or both and often become intimately intertwined with other design decisions.
To distinguish between different kinds of design decisions we classify them into three categories: (1) simple, (2) compound, and (3) crosscutting. Simple decisions have a singular rationale and consequence. Compound decisions include several closely related rationales, but their consequences are generally contained to one component. Finally, crosscutting decisions affect a wider range of components and their rationale follows a higher level concern such as architectural quality of the system.

\subsection{Architecture Recovery, Change, and Decay Evaluator}

In order to capture the consequence aspect of design decisions, we build on top of the existing work in architecture modeling and recovering.
To obtain the static architectures of a system from its source code, we use our recent work which resulted in a novel approach, called \ARCADE~\cite{behnamghader2016large,le2015empirical}. \ARCADE is a software workbench that employs (1) a suite of architecture-recovery techniques and (2) a set of metrics for measuring different aspects of architectural change. It constructs an expansive view showcasing the actual (as opposed to idealized) evolution of a software system's architecture. 

\ARCADE  allows an architect (1) to extract multiple architectural views from a system's codebase and (2) to study the architectural changes during the system's evolution as reflected in those views. 
\ARCADE currently provides access to ten recovery techniques. 
We use two of these techniques in this paper. These two techniques are \emph{Algorithm for Comprehension-Driven Clustering} (\ACDC)~\cite{tzerpos2000acdc} and \emph{Architecture Recovery using Concerns} (\ARC)~\cite{garcia2011enhancing}. 
Our previous evaluation~\cite{garcia2013comparative} showed that these two techniques exhibit the best accuracy and scalability of the ten.
\ACDC's view is oriented toward components that are based on structural patterns (e.g., components consisting of entities that together form a particular subgraph). On the other hand, \ARC's view produces components that are semantically coherent due to sharing similar system-level concerns (e.g., a component whose main concern is handling distributed jobs).

To measure architectural changes across the development history of a software system,  \arcade provides several architecture similarity metrics: cvg \cite{le2015empirical} and a2a \cite{behnamghader2016large}, MoJo \cite{tzerpos1999mojo}, and MoJoFM \cite{wen2004effectiveness}. These are system-level similarity metrics calculated based on the cost of transforming one architecture to another. Using similar principles, \ourappr conducts the change analysis and extracts a system's architectural changes (refer to Section \ref{sec:ca}).

\section{Approach}\label{sec:approach}
Knowledge vaporization in software systems plays a major role in increasing maintenance costs, and  exacerbates architectural drift and erosion \cite{jansen2005software}. The goal of \ourappr is to uncover architectural design decisions in software systems, and thereby help reverse the course of knowledge vaporization by providing a crisper understanding of such decisions and their effects. 

This section aims at providing answers to the five questions raised in the introduction. We elaborate on the \mbox{\emph{definition}} and \mbox{\emph{classification}} of design decisions. 
We describe how architectural changes that are the \emph{reifications} of architectural design decisions can be recovered from the source code of real software systems.  
We also describe a process whereby architectural  decisions are \emph{identified} in real systems. Finally, our approach enables engineers to continuously capture  architectural decisions in software systems during their \mbox{\emph{evolution}}.

In Section \ref{sec:f-d}, we identified two constituent parts of an architectural design decision, \textit{rationale} and \textit{consequence}. 
The static architecture of a system  explicitly captures the system's components and possibly other architectural entities, but rationale is usually missing or, at best, implicit in the structural elements. 
For this reason, our approach focuses on the consequences of design decisions. We have developed a technique that leverages the combination of source code and issue repositories to obtain the design decision consequences. 
Issue repositories are used to keep of track of bugs, development tasks, and feature requests in a software development project. Code repositories contain historical data about the inception, evolution, and incremental code changes in a software system. Together, these repositories provide the most reliable and accurate pieces of information about a  system.

\ourappr 
automatically extracts the required information from a system's repositories and outputs the set of design decisions made during the system's lifecycle. In order to achieve this \ourappr first recovers the static architecture of the target system. \ourappr then cross-links the issues to their corresponding code-level changes. These links are in turn analyzed to identify candidate architectural changes and, subsequently, their rationales.

\begin{figure}[t]
	\centering
	\includegraphics[width = .32\textwidth]{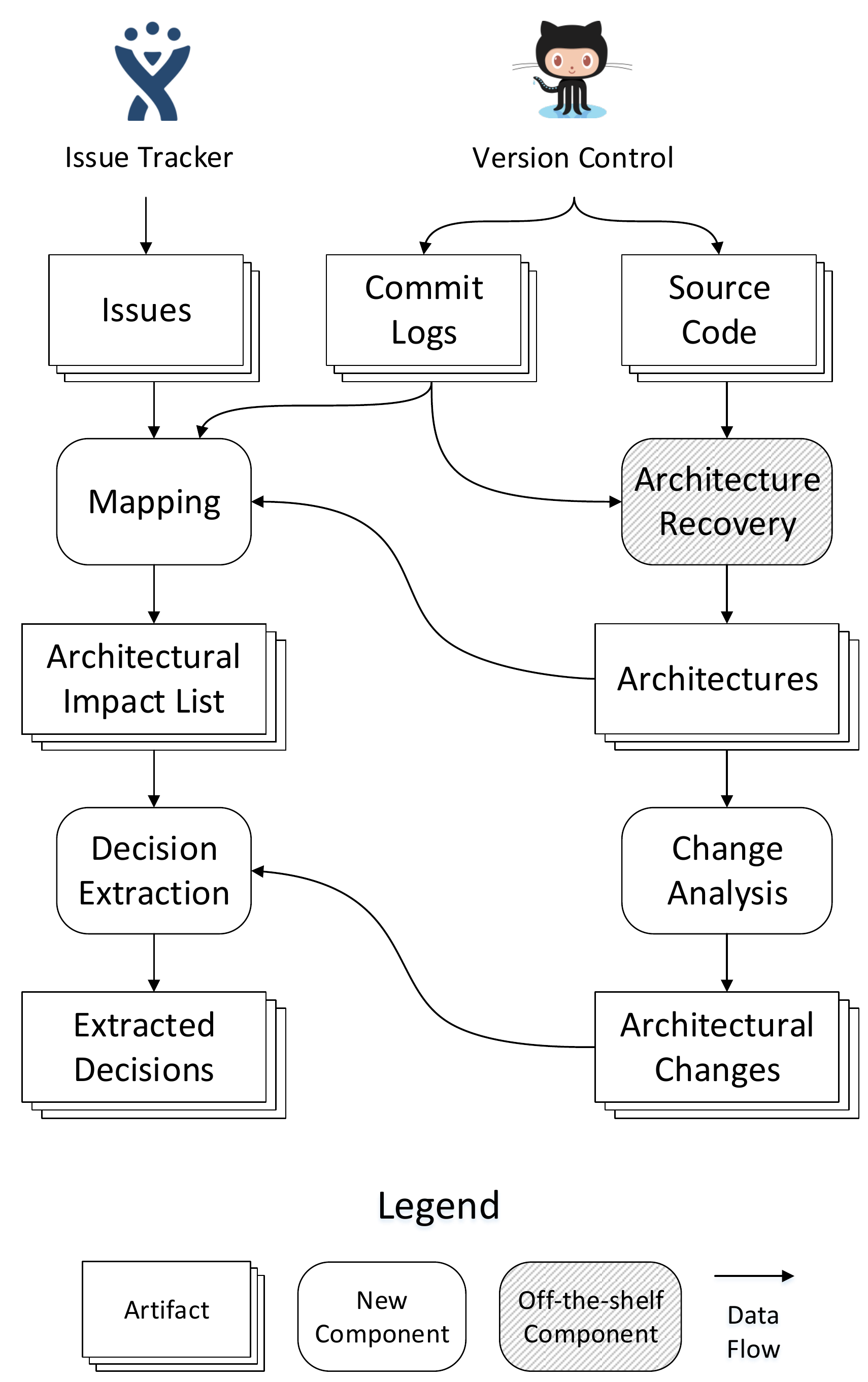}
	\caption[Overview of ]{Overview of \ourappr. 
		Using the existing source code, commit logs, and extracted issues obtained from code and issue repositories, our approach automatically extracts the underlying design decisions. Implementation of the new components spans over 4,000 Source Lines of Code (SLoC).}
	\label{fig:overview}
\end{figure}

A high level overview of \ourappr's workflow is displayed in Figure \ref{fig:overview}.  \ourappr begins by recovering the static architecture of a system. This step is only required if an up-to-date, reliable, documented architecture is not available. 
After recovering or obtaining the architectures of different versions of its target system, \ourappr follows through three distinct phases. In the first phase (\textbf{Change Analysis}) \ourappr identifies how the architecture of the system has changed along its evolution path. The second phase (\textbf{Mapping}) mines the system's issue repository and creates a mapping (called \textit{architectural impact list}) from issues to the architectural entities they have affected. Finally, the third phase (\textbf{Decision Extraction}) creates an overarching decision graph by putting together the architectural changes and the architectural impact list. This graph is in turn traversed to uncover the individual design decisions. In the remainder of this section we detail each of the three phases. 

\subsection{Change Analysis}\label{sec:ca}

Architectural change has been recognized as a critical phenomenon from the very beginnings of the study of software architecture \cite{perry1992foundations}. However, only recently have researchers tried to empirically measure and analyze architectural change in software systems \cite{le2015empirical, behnamghader2016large}. These efforts rely on architectural change metrics that quantify the extent to which the architecture of a software system changes as the system evolves. This work has served as a motivation and useful foundation in obtaining a concrete view of architectural changes. 

Specifically, we have designed \emph{Change Analyzer} (\emph{CA}), which is inspired by the manner in which existing metrics (e.g., a2a \cite{behnamghader2016large}, MoJo \cite{tzerpos1999mojo}, and MoJoFM \cite{wen2004effectiveness}) measure architectural change. 
These metrics consider five operations used to transform architecture A into architecture B: addition, removal, and relocation of system entities (e.g., methods, classes, libraries) from one component to another, as well as addition and removal of components themselves \cite{agnew1994planning, medvidovic1996, oreizy1998}.
We use a similar notion and define architectural change as a set of \textit{architectural deltas}. An architectural delta is: (1)~any modification  to a component's internal entities including additions and removals (a relocation is treated as a combined addition to the destination component and removal from the source component), or (2)~additions and removals of entire components.
We then aggregate these deltas into architectural change instances. 
Algorithm \ref{alg:ca} describes the details of the approach used to extract the architectural deltas and changes. 

\emph{CA} works in two passes. In the first pass, \emph{CA} matches the most similar components in the  given pair of architectures. In the second pass, \emph{CA} compares the matched components,  extracts the architectural delta(s), and clusters them into architectural change instances. 

The objective of the matching pass is to find the most similar components in a way that minimizes the overall difference between the matched components. Since two architectures can have different numbers of components, \emph{CA} first balances (Algorithm \ref{alg:ca}, line \ref{alg:ca:balance}) the two architectures. To do so, \emph{CA} adds ``dummy'' (i.e., empty) components to the architecture with  fewer components until both architectures have the same number of components. After balancing the architectures, \emph{CA} creates a weighted bipartite graph from architecture A to architecture B and calculates the cost of each edge. Existence of an edge denotes that component $C_A$ has been transformed into component $C_B$. The cost of an edge is the total number of architectural deltas required to effect the transformation. 

\begin{algorithm}[t]
	\renewcommand{\AlCapSty}[1]{\small\small{\textbf{#1}}\unskip}
	\algsmall
	\label{alg:ca}
	\SetAlgoVlined
	\LinesNumbered 
	\DontPrintSemicolon
	\SetInd{0.3em}{0.6em}
	
	\KwIn{$Architecture A, Architecture B$}
	\KwOut{$Changes \Leftarrow$ a set of architectural changes}
	Let $ComponentsA$ = $ArchitectureA$'s components \\
	Let $ComponentsB$ = $ArchitectureB$'s components \\
	Let $E_{all}$, $E_{chosen}$ = $\emptyset$  \\
	
	\If {$|ComponentsA| \neq |ComponentsB|$}{
			$Balance(ComponentsA, ComponentsB)$ \label{alg:ca:balance}
		}
	\ForEach{  $c_a$ $\in \emph{ComponentsA}$  } {    
		\ForEach{  $c_b$ $\in \emph{ComponentsB}$  } {    
				$cost = CalculateChangeCost(c_a, c_b)$ \\
				$e = \{c_a, c_b, cost\}$ \\
				$add~e~to~E_{all}$
			}
	}
	
	$E_{chosen}= MinCostMatcher(ComponentsA, ComponentsB, E_{all})$\label{alg:ca:mincost}\\
	\ForEach{$e \in E_{chosen}$}{
		$Changes = GetChangeInstances(e.c_a, e.c_b) \cup Changes$
		}
	$return ~Changes$
	
	\caption{Change Analysis}
\end{algorithm}
\begin{algorithm}[h]
	\renewcommand{\AlCapSty}[1]{\small\small{\textbf{#1}}\unskip}
	\algsmall
	\label{alg:gad}
	\SetAlgoVlined
	\LinesNumbered 
	\DontPrintSemicolon
	\SetInd{0.3em}{0.6em}
	
	\KwIn{$ComponentA, ComponentB$}
	\KwOut{$Changes$}
	Let $EntitiesA$ =  $ComponentA$'s entities \\
	Let $EntitiesB$ =  $ComponentB$'s entities \\

		\If {$EntitiesA \cap EntitiesB = \emptyset$}{
			$Change ~ch_1,ch_2$\\
			$ch_1.deltas = EntitiesA$\\
			$ch_2.deltas = EntitiesB$\\
			$return ~\{ch_1, ch_2\}$
		}
		\Else {
			$Change ~ch$\\
			$ch.deltas = (EntitiesA \cup EntitiesB)  - (EntitiesA \cap EntitiesB)$\\
			
			$return ~\{ch\}$\\
		}
	
	\caption{$GetChangeInstances$ method}
\end{algorithm}

\begin{figure}[b]
	\centering
	\includegraphics[width = \columnwidth]{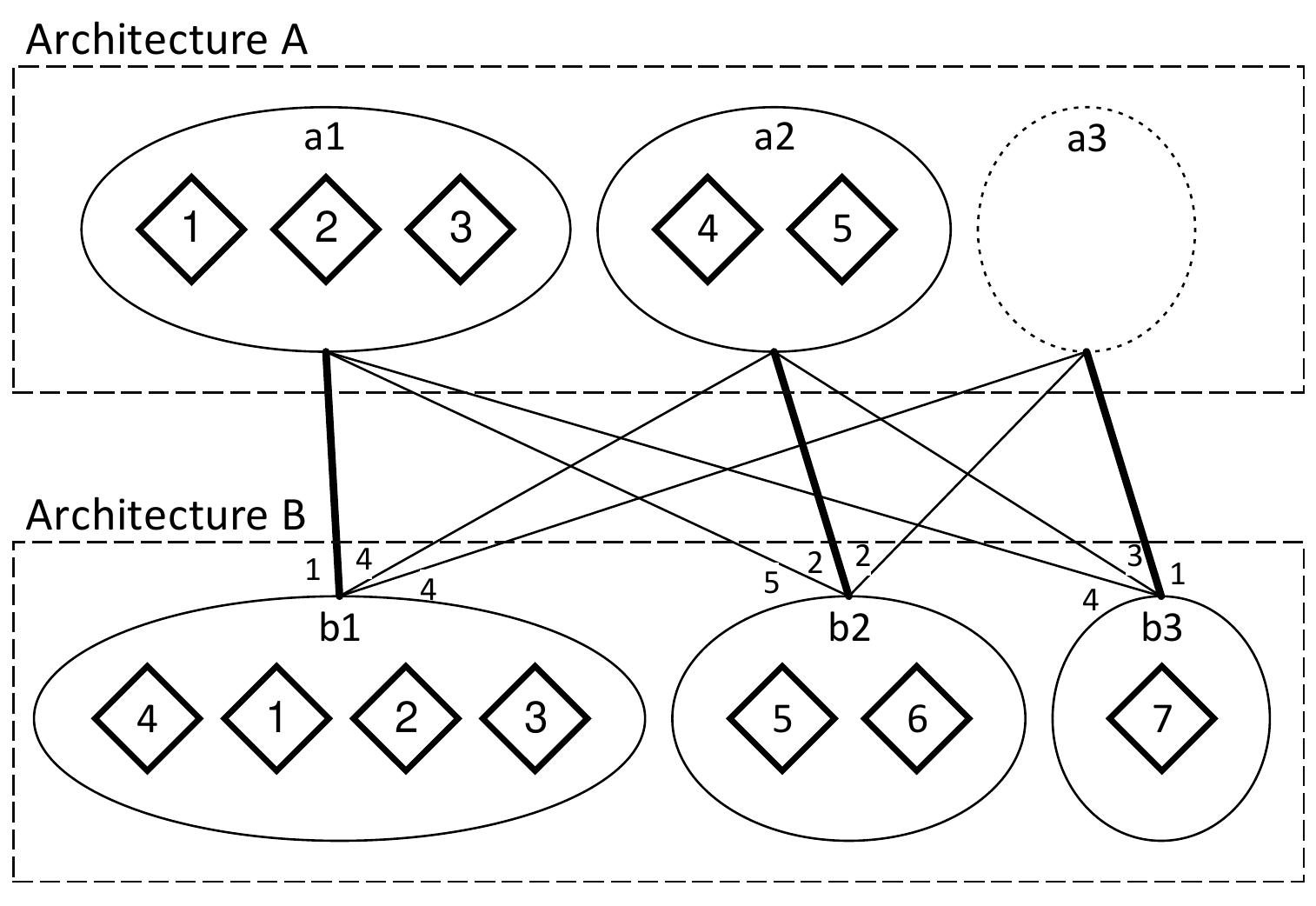}
	\caption[]{Calculating the costs of the edges and finding the perfect matching. The bold connectors are the selected edges that lead to minimum overall cost.}
	\label{fig:change_matching}
\end{figure}

Figure \ref{fig:change_matching} displays a simple example of two architectures and the corresponding bipartite graph with all  possible edges. \textit{MinCostMatcher} (Algorithm \ref{alg:ca}, line \ref{alg:ca:mincost}) takes the two architectures and the set of edges between them, and selects the edges in a way that ensures a bijective matching between the components of the two architectures with the minimum overall cost (sum of the costs of the selected edges). 
\emph{MinCostMatcher} is based on linear programming; its details are omitted for brevity.

\begin{figure}[t]
\vspace{-2mm}
	\centering
	\includegraphics[width = \columnwidth]{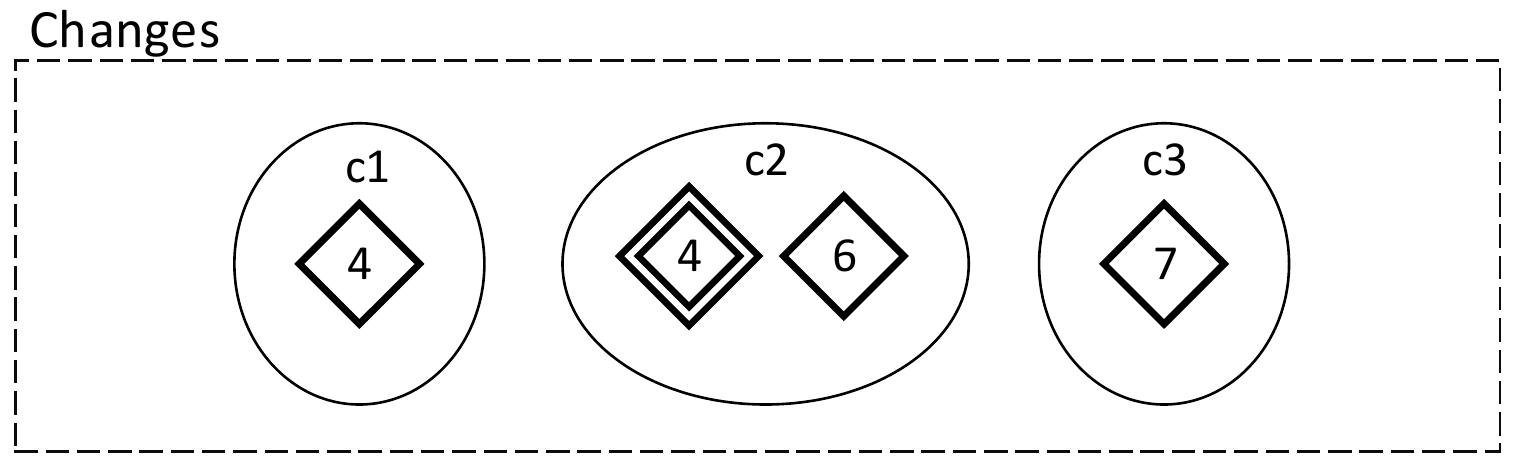}
	\caption[]{Extracted changes between the architectures depicted in Fig. \ref{fig:change_matching}. Double-lined diamonds indicate removals while regular diamonds denote additions.}
	\label{fig:changes}
\end{figure}

In the second pass, \emph{CA} extracts the architectural deltas between the matched components. If there are no common architectural entities between two matched components, we create two change instances, one for the component that has been removed and one for the newly added component. The reason is to distinguish between  transformations of components and their additions and removals. Figure \ref{fig:changes} depicts the extracted changes of our example architectures.

\subsection{Mapping}
The output of \emph{CA} is a set of architectural changes that is a superset of the consequences of  design decisions. 
The goal of \emph{Mapping} is to find all the issues that point to  the rationale  of the design decisions that yielded those consequences.
To that end, \emph{Mapping} first identifies the issues that satisfy two conditions: (1)~they belong to the version of the system being analyzed and (2)~they have been marked as resolved and their consequent code changes have been merged with the main code base of the system. \emph{Mapping} then extracts the code-level entities affected by each issue. These code-level entities are identified by mining the issues' commit logs and pull requests. 
%
Using one or more architecture recovery methods available in \arcade, the code-level entities are translated into corresponding architectural entities. The list of all issues, as well as the mapping between the issues and the architectural entities affected by them is called the \emph{Architectural Impact List}.

\begin{figure}[b]
	\centering
	\includegraphics[width = .7\columnwidth]{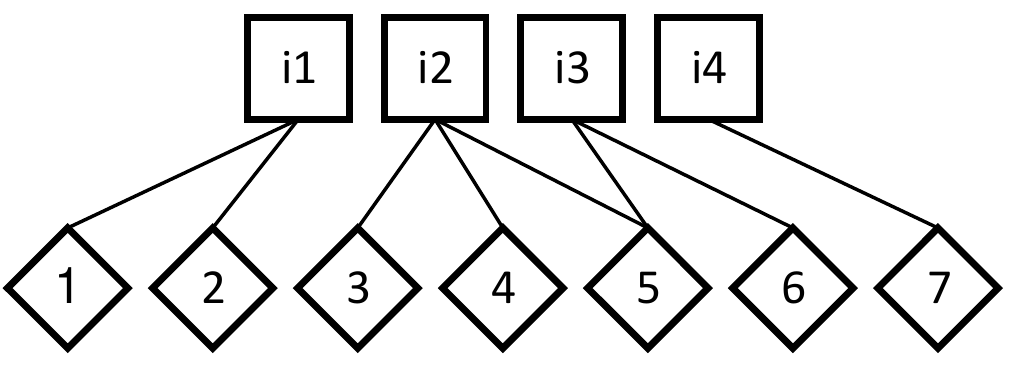}
	\caption[]{Architectural impact list. Squares represents issues and diamonds represent entities. An edge from an issue to an entity means that resolving that issue resulted in modifying that entity.}
	\label{fig:impact}
\end{figure}

A graph-based view of this list is displayed in Figure \ref{fig:impact}. It is possible for issues to have overlapping entities (e.g., i2 and i3 are both connected to entity 5).  It is also important to note that the presence of an edge from an issue to an entity does not necessarily indicate architectural change (e.g., entities 1 and 5 are not part of any of the architectural changes displayed in Figure \ref{fig:changes}). This is intuitively expected, since a great many of issues do not incur substantial enough change in the source code and thereby the architecture of the system.

\subsection{Decision Extraction}\label{sec:decisionanalysis}
In its final phase, \ourappr creates the overarching decision graph by putting together the architectural changes and their pertaining issues. This graph is traversed and  individual design decisions are identified. Algorithm \ref{alg:decision} details this phase.

\begin{algorithm}[t]
	\renewcommand{\AlCapSty}[1]{\small\small{\textbf{#1}}\unskip}
	\algsmall
	\label{alg:decision}
	\SetAlgoVlined
	\LinesNumbered 
	\DontPrintSemicolon
	\SetInd{0.3em}{0.6em}
	
	\KwIn{$ArchitecturalImpactList, Changes$}
	\KwOut{$Decisions$}
	Let $DecisionsGraph$ = bipartite graph of decisions \\
	\ForEach{  $(issue, entities) \in ArchitecturalImpactList$  } {    
		\ForEach{  $c$ $\in Changes$  } {    
				\If {$c.deltas \cap entities \neq \emptyset$}{
					$connect (issue, c) ~in ~DecisionsGraph$
				}
		}
	}
	$Decisions = FindDecisions(DecisionsGraph)$\\
	$return ~~Decisions$

	\caption{$Decision ~Extraction$}
\end{algorithm}
\begin{figure}[b]
	\centering
	\includegraphics[width = 0.4\columnwidth]{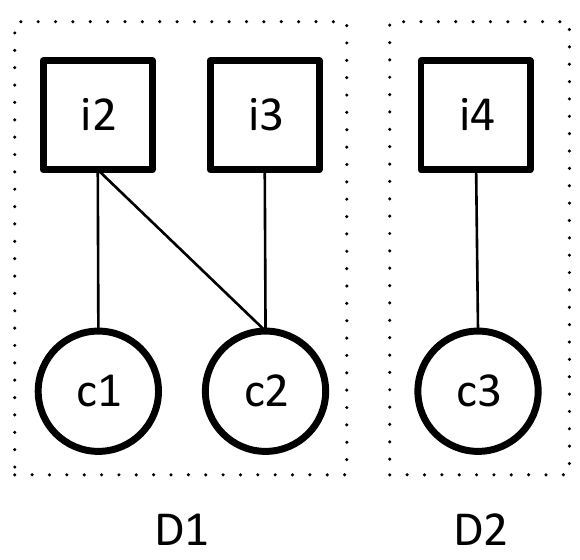}
	\caption[Decisions]{The overarching decisions graph contains two decisions D1 and D2. Squares denote issues, and circles denote changes.  }
	\label{fig:decisions}
\end{figure}

Algorithm \ref{alg:decision} traverses the architectural impact list generated in the \emph{Mapping} phase and the list of changes. If there is an intersection between the entities matched to issues and the entities involved in changes, then it adds an edge connecting the issue with the change. 
The intuition behind this is that an issue contains the rationale for a decision if it affects the change(s), which are the consequences, of that decision. 
We note that, hypothetically, there can be situations in which an issue is the cause of a change without directly affecting any architectural deltas in that change.
For example, if an issue leads to removing all the dependencies to an entity, that entity might get relocated out of its containing component by the architecture recovery technique. 
However, detecting these situations in a system's architecture is not possible with existing recovery techniques, because they abstract away the dependencies among internal entities of a component.  Although such information could easily be incorporated, \ourappr would be unable to deal with such scenarios as currently implemented.

The decisions graph for our running example is depicted in Figure \ref{fig:decisions}. The
$FindDecisions$ method in  Algorithm \ref{alg:decision} removes all  orphaned changes and issues, and in the remaining graph locates the largest disconnected subgraphs. Each disconnected subgraph represents a decision. The reason is that these disconnected subgraphs are the largest sets of interrelated rationales and consequences that do not depend on other issues or changes. Intuitively, we expect that, in a real-world system, only a subset of issues will impose changes whose impact on the system can be considered  architectural. Furthermore, each of those issues will reflect a specific, targeted objective. Therefore, in a typical system, the graph of changes and issues  should contain disconnected subgraphs of reasonable sizes. This is discussed further in our evaluation in Section \ref{sec:evaluation}.  

In Section \ref{sec:foundation}, we identified three different types of decisions: (1) Simple decisions are the decisions that consist of a single change and a single issue. These decisions have a clear rationale and consequence. (2) Compound decisions are the decisions that include multiple issues and a single change. These decisions are similar to simple decisions and the issues involved are closely related to an overarching rationale.  Finally, (3) cross-cutting decisions are the decisions that include multiple changes and one or more issues. These decisions have a higher-level, ``compound'' rationale---e.g., improving the reliability or performance of a system---that requires multiple changes to be achieved. In Figure \ref{fig:decisions}, D$_1$ is a cross-cutting decision while D$_2$ is a simple decision.

For illustration, Table \ref{tab:decisions} lists three real examples of  decisions, one of each type, uncovered from Hadoop. Information in the \emph{Issue(s)} column 
contains the summaries of the issues pertaining to that decision. Each boxed number indicates a separate issue or change. The data in the \emph{Change(s)} column is a short description of the changes involved in a given decision.
The simple decision in the top row is an update to satisfy a requirement by changing the \emph{job tracking module}. The compound decision in the middle row describes the two sides of a problem that is resolved by changing the \emph{compression module} of Hadoop. Finally the uncovered crosscutting decision in the bottom row is about a series of changes applied to increase the reliability of Hadoop's task execution by preventing data corruption, and providing two facilities that make it easier to check the status of the tasks that are being executed. 


\begin{table}[]
	\centering
	\caption{Examples of uncovered decisions in Hadoop}
	\label{tab:decisions}
	\begin{tabular}{p{1.3cm} p{3.5cm} p{3cm}}
		\toprule
		\textbf{Decision Type}        & \textbf{Issue(s)}                                                                                                                                                                                & \textbf{Change(s)}                                                      \\ \midrule
		Simple                        & \boxed{1} Job tracking module only kept track of the jobs executed in the past 24 hours. If and admin checked the history after a day of inactivity, e.g., on Monday, the list would be empty. & \circled{1}  \texttt{hadoop.mapred} component was modified  \\ \midrule
		\multirow{2}{*}{Compound}     & \boxed{1} UTF8 compressor does not handle end of line correctly                                                                                                                                & \circled{1} \texttt{CompressionInputSt{\allowbreak}ream} was added                \\
		& \boxed{2} Sequenced files should support `custom compressors'                                                                                                                                  & and \texttt{CompressionCodec} was modified.                              \\ \midrule
		\multirow{3}{*}{Crosscutting} & \boxed{1} Random seeks corrupts \texttt{InputStream} data                                                                                                                                      & \circled{1} \texttt{hadoop.streaming} was modified                                     \\
		& \boxed{2} Streaming should send status signals  every 10 seconds                                                                                                                & \circled{2} \texttt{hadoop.metrics} was modified                                   \\
		& \boxed{3}  Task status should include timestamp for job transitions                                                                                                                            & \circled{3} \texttt{hadoop.fs} was modified                                            \\ \bottomrule

	\end{tabular}
\end{table}


 




\section{Evaluation}\label{sec:evaluation}
We have empirically evaluated \ourappr to verify its applicability and measure its accuracy in uncovering architectural design decisions. Section IV-A discusses the real-world systems on which \ourappr was applied, demonstrating its applicability. Sections IV-B and IV-C discuss the precision and recall of our results, respectively, demonstrating \ourappr's accuracy.

\subsection{Applicability}
Table \ref{tab:systems} contains information about the two subject systems we have used in our evaluation.
These systems were selected from the catalogue of Apache open-source software systems~\cite{apache}. 
We selected Hadoop \cite{hadoop:homepage} and Struts \cite{struts} because they are widely adopted and fit the target profile of  candidate systems for our approach: they are open-source, and have accessible issue and code repositories. 
Furthermore, these systems are at the higher end of the Apache software systems' spectrum in terms of size and lifespan. Both of these projects use GitHub \cite{github} as their version control and source repository, and Jira \cite{jira} as their issue repository. 
We analyzed more than 100 versions of Hadoop and Struts in total. Our analyses spanned over 8 years of development, 35 million SLoC, and over 4,000 resolved issues. 

\begin{table}[]
	\centering
	\caption{Subject systems analyzed in our study}
	\label{tab:systems}
	\begin{tabular}{@{}lllll@{}}
		\toprule
		\textbf{System} & \textbf{Domain}        & \textbf{No. Ver.} & \textbf{No. Iss.} & \textbf{MSLoC} \\ \midrule
		Hadoop          & Distributed Processing & 68                   & 2969                 & 30.0           \\
		Struts          & Web Apps Framework     & 36                   & 1351                 & 6.7            \\ \bottomrule
	\end{tabular}
\end{table}

An overview of the results of applying \ourappr to the two subject systems is depicted in Table \ref{tab:summary}. These results are grouped by (1) system (Hadoop vs. Struts) and (2) employed recovery technique (\arc vs. \acdc).
In this table, \emph{No. of Iss. in Decisions} represents the total number of issues that were identified to be part of an architectural design decision. On average, only about 18\% of the issues for Hadoop and 6\% of the issues for Struts have had architecturally significant effects, and hence have been considered  parts of a design decision. 
This is in line with the intuition that only a subset of issues will impose changes whose impact on the system can be considered  architectural. 
Morevoer, this observation bolsters the importance of \ourappr for understanding the current state of a system and the decisions that have led to it. Without having access to \ourappr, architects would have to analyze 5-to-15 times more issues and commits to uncover the rationales and root causes behind the architectural changes of their system. 
The remainder of  Table \ref{tab:summary} displays the total number of detected architectural changes (\emph{No. of Changes}), the total number of uncovered architectural design decisions (\emph{No. of Decisions}), and the average numbers of issues and changes per decision (\emph{Avg. Issues/Decision} and \emph{Avg. Changes/Decision}, respectively). 
It is worth mentioning that not all the detected changes were matched to design decisions. We will elaborate further on this in Section \ref{sec:recall}, which evaluates \ourappr's recall.

As displayed in Table \ref{tab:summary}, depending  on the  technique used to recover the architecture, the number of uncovered design decisions varies. The reason is that \acdc and \arc approach architecture recovery from different perspectives: \acdc leverages a system's module dependencies; \arc derives a more semantic of a system's architecture using natural language processing and information retrieval techniques. Therefore, the nature of the recovered architectures and changes, and consequently the uncovered design decisions, are different. In our previous work, we showed that these recovery techniques provide complementary views of a system's architecture \cite{le2015empirical}. 
The propagation of these complementary views to our approach has yielded some tangible effects. 
For instance, \ourappr running \arc was able to uncover a decision 
about refactoring the names of a set of classes and methods in Hadoop, 
while \ourappr running \acdc could not uncover that decision. 
The reason is that \arc  is sensitive to lexical changes by design. 
Depending on the context and objectives of using \ourappr, architects can choose the most favorable view for their purposes.

\begin{figure}[b]
 	\vspace{-2mm}
 	\centerline{
 		\subfigure[Hadoop-ACDC]{\includegraphics[width=.24\columnwidth]{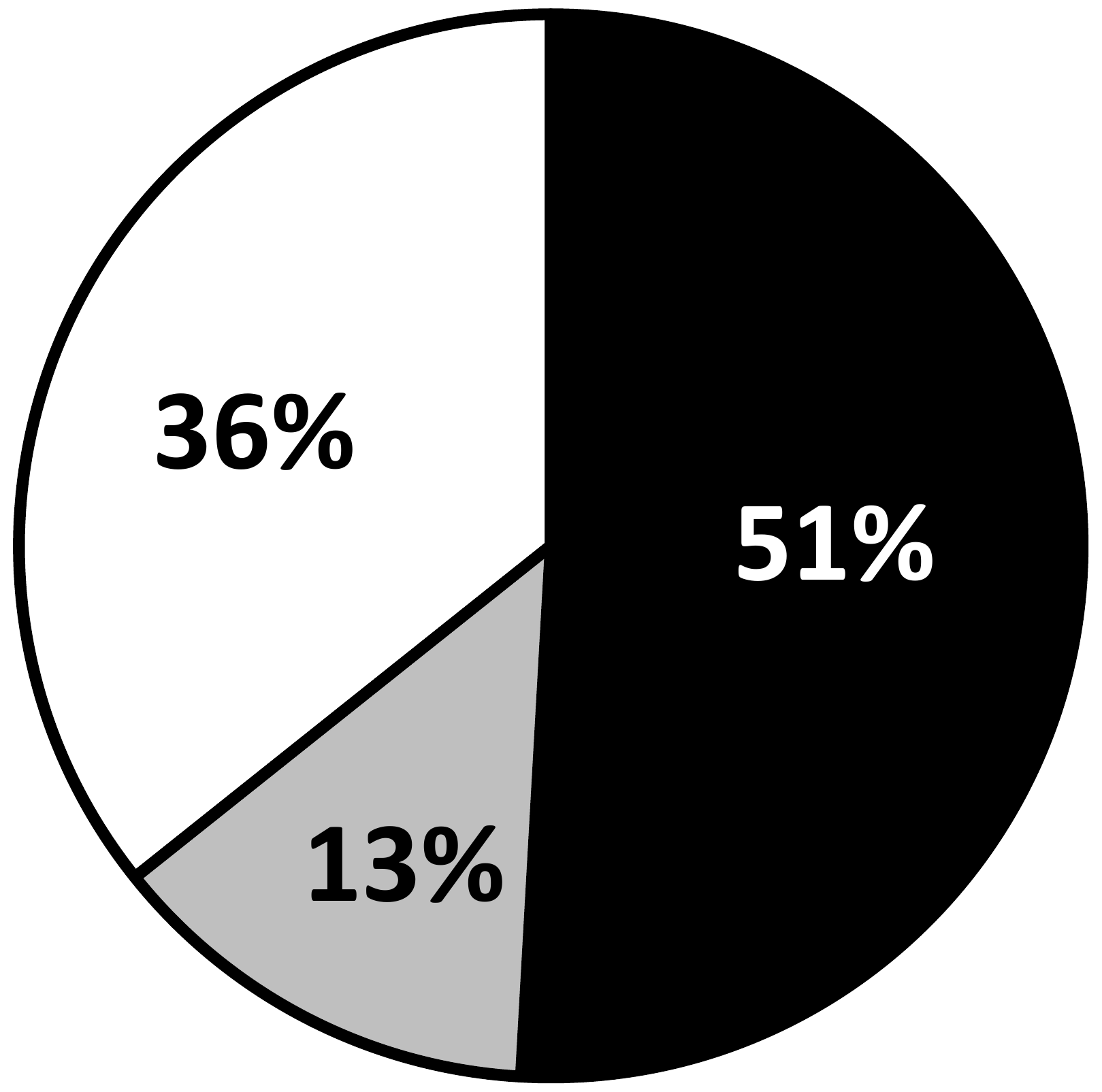}\label{fig:hadoopacdc}}
 		\subfigure[Hadoop-ARC]{\includegraphics[width=.24\columnwidth]{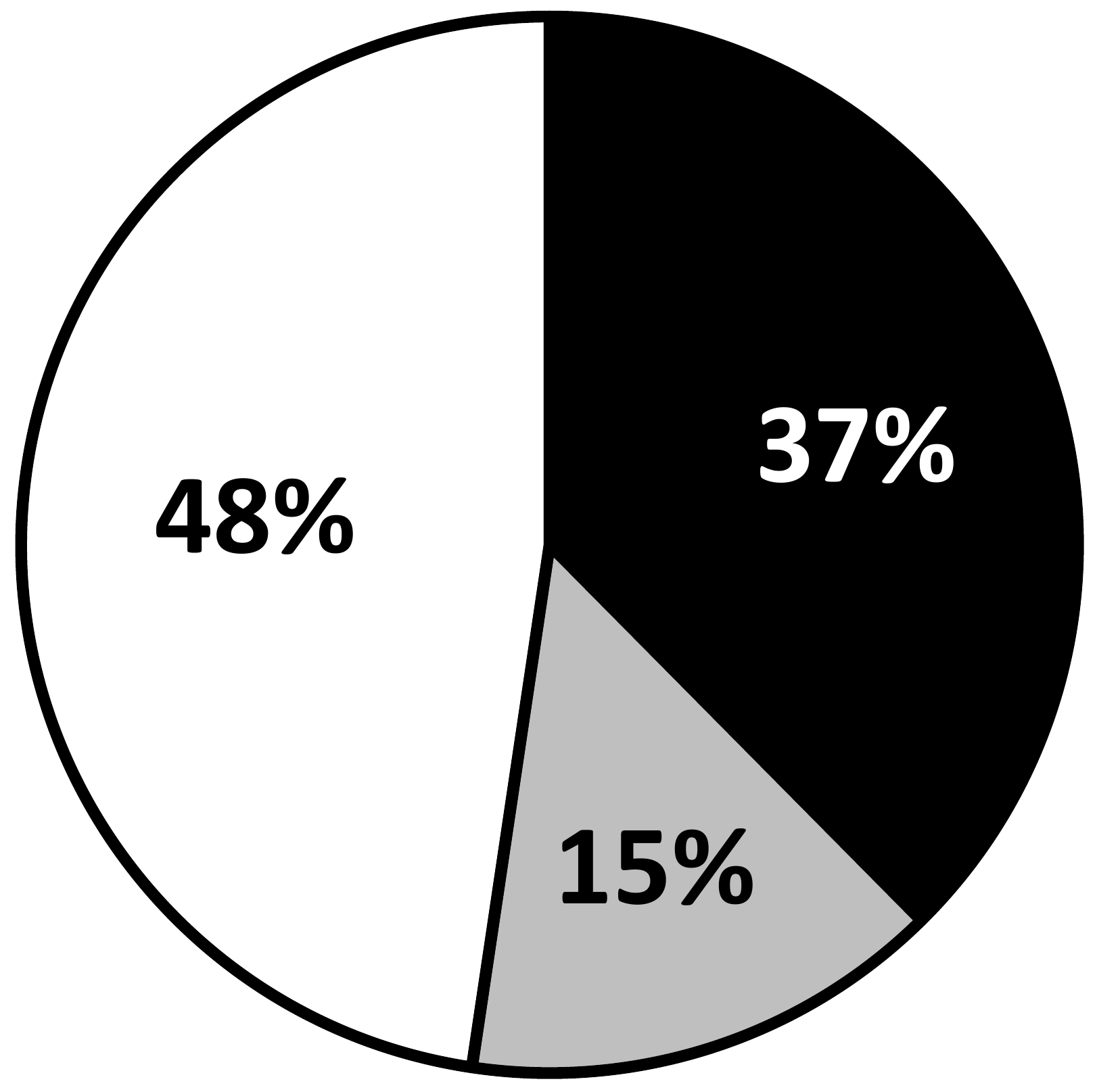}\label{fig:hadooparc}}
		\subfigure[Struts-ACDC]{\includegraphics[width=.24\columnwidth]{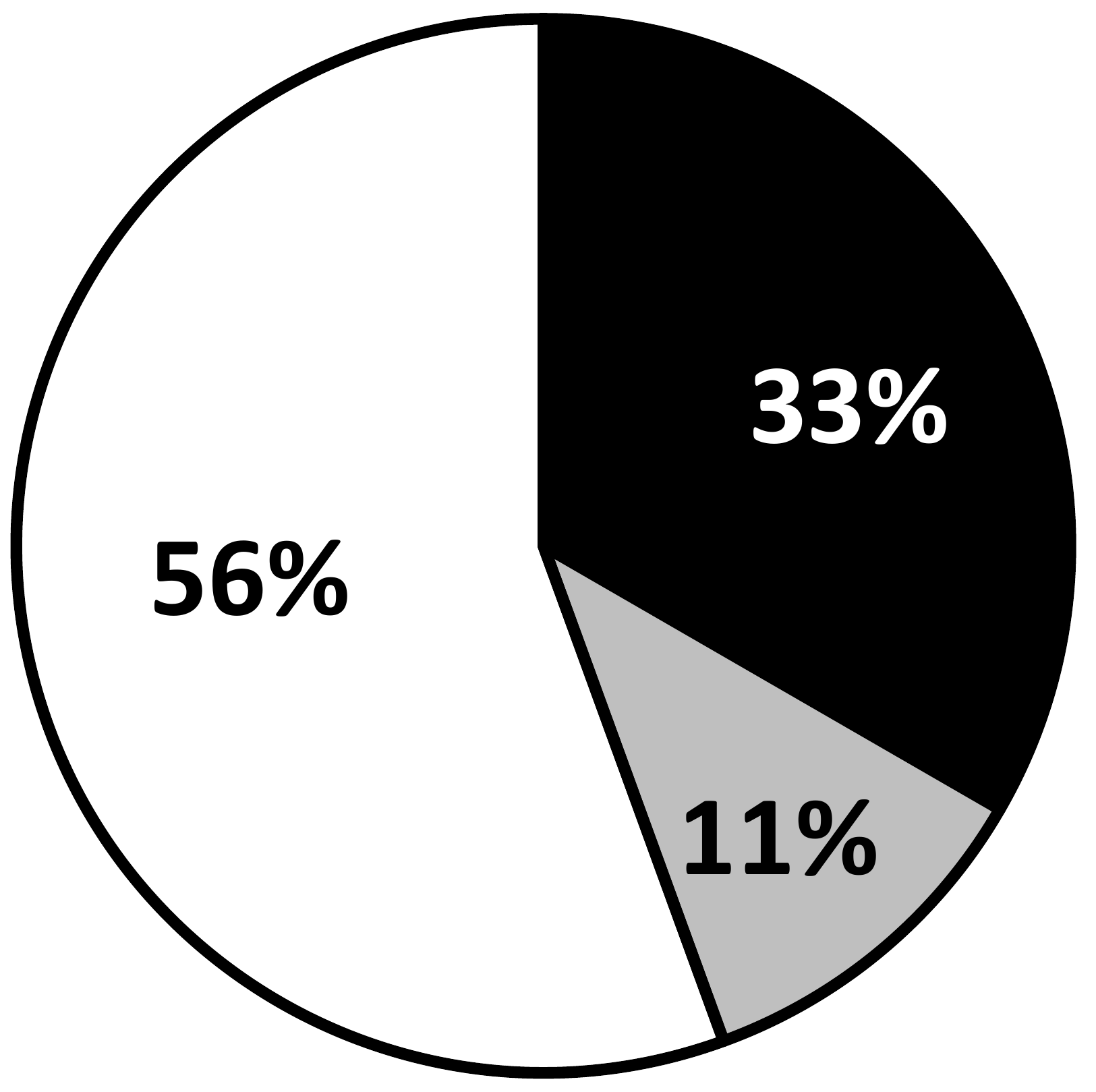}\label{fig:strutsacdc}} 	
		\subfigure[Struts-ARC]{\includegraphics[width=.24\columnwidth]{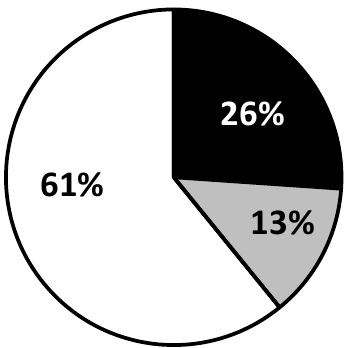}\label{fig:strutsarc}}}

 	\caption{Distribution of different types of decision in the subject systems: solid black  represents simple decisions; grey  denotes compound decisions;  white  displays cross-cutting decisions.}
 	\label{fig:distribution}
 \end{figure}

\ourappr aims to uncover three kinds of architectural design decisions (recall Section \ref{sec:foundation}). Our results confirmed the presence of all three kinds in our subject systems. Figure \ref{fig:distribution} depicts the distribution of different kinds of decisions detected for each pair of systems and recovery techniques. While the relative proportion of simple and cross-cutting decisions varies across systems and employed recovery techniques, the number of compound decisions is consistently the smallest. 
%
	One possible explanation is that as a system matures resolution of its issues either become more isolated to a smaller scope that leads to detection of simple decisions, or more quality drivern that leads to detection of crosscutting decisions.

\begin{table}[]
	\centering
	\caption{Overview of the results}
	\label{tab:summary}
	\begin{tabular}{@{}lcccc@{}}
		\toprule
		\multirow{2}{*}{\textbf{Systems}} & \multicolumn{2}{c}{\textbf{Hadoop}} & \multicolumn{2}{c}{\textbf{Struts}} \\ \cmidrule(l){2-5} 
		                                  & \textbf{ACDC}     & \textbf{ARC}    & \textbf{ACDC} & \textbf{ARC}\\ \midrule
		\textbf{No. of Iss. in Decisions}  & 427               & 674             & 70            & 94        \\
		\textbf{No. of Changes}           & 950               & 3935            & 220           & 1359      \\
		\textbf{No. of Decisions}         & 112               & 149             & 27            & 23        \\
		\textbf{Avg. Issues/Decision}     & 3.81              & 4.52            & 2.59          & 4.94      \\
		\textbf{Avg. Changes/Decision}    & 1.77              & 2.36            & 1.77          & 2.21      \\
		\bottomrule
	\end{tabular}
\end{table}
\subsection{Precision}
In order to assess the precision of \ourappr, we need to determine whether the uncovered architectural design decisions are valid. As captured in the premise of \ourappr, architectural design decisions are not generally documented, hence a ``ground-truth'' for our analyses was not readily available.

To overcome this hurdle, we devised a systematic plan to objectively assess the rationales and consequences of the uncovered design decisions. We defined a set of criteria targeting the two aspects of an architectural design decision---rationale and consequence---and used them as the basis of our assessment. 
Two PhD students carefully carried out the analysis and the results of their independent examinations were later aggregated. 
In the remainder of this section, we will elaborate on the details of the conducted analyses.


We use four criteria targeting different parts of an architectural design decision (two targeting rationales and two targeting consequences). Each criterion is rated using a three-level-scale, with the numeric values of 0, 0.5, and 1. In this scale, 0 means that the criterion is not satisfied; 0.5 means that the satisfaction of the criterion is confirmed after further investigation by examining the source-code, details of the issues, or commit logs; finally, 1 means that the criterion is evidently satisfied.
The reason we use a three-level scale in our analysis is to measure the precision of the results of our approach from the viewpoint of non-experts, and to distinguish the decisions according to the effort required for understanding them. 
To that end, any criterion whose evaluation requires (1) in-depth system expertise, (2) inspection of information other than that captured in design decisions, and/or (3) having access to the original architects of the system, is given a rating of 0.

The criteria for assessing rationales are two-fold:
\begin{enumerate}
\item \emph{Rationale Clarity} indicates whether the rationale and its constituent parts are clear and easily understandable. This is accomplished by looking at the summaries of the issues and pinpointing the problems or requirements driving the decision.
\item \emph{Rationale Cohesion} indicates the degree to which there is a coherent relationship among   the issues that make up a given rationale.  
\emph{Rationale Cohesion} is only analyzed if the decision is shown to possess the \emph{Rationale Clarity} criterion.
\end{enumerate} 

The criteria for assessing consequences are also two-fold:
\begin{enumerate}
\item \emph{Consequence-Rationale Association} assesses whether the changes and their constituent architectural deltas are related to the listed rationale.
\item \emph{Consequence Tractability} assesses whether the size of the changes is tractable. In other words, is the number of changes and their constituent deltas small enough to be understandable in a short amount of time?\footnote{As a rule of thumb, decisions including more than five changes did not satisfy this criterion in our evaluation. This is, of course, adjustable.}
\emph{Consequence Tractability} is only analyzed if the decision is shown to possess the \emph{Rationale Clarity} criterion.
\end{enumerate} 


\begin{table}[b]
	\centering
	\caption{Average Decision Scores}
	\label{tab:averages}
	\begin{tabular}{@{}lcccc@{}}
		\toprule
		\multirow{2}{*}{\textbf{Decisions}} & \multicolumn{2}{c}{\textbf{Hadoop}} & \multicolumn{2}{c}{\textbf{Struts}} \\ \cmidrule(l){2-5} 
		& \textbf{\acdc}     & \textbf{\arc}    & \textbf{\acdc}     & \textbf{\arc}    \\ \midrule
		\textbf{Simple}                     & 0.89           &  0.95    & 0.90              &    0.99         \\
		\textbf{Compound}                   & 0.50           &  0.52    & 0.76              &    0.56         \\
		\textbf{Crosscutting}               & 0.61           &  0.76    & 0.78              &    0.77         \\
		\textbf{Overall}                    & 0.72           &  0.72    & 0.81              &    0.71         \\ \bottomrule
	\end{tabular}
\end{table}

The two PhD students independently scored every decision based the above criteria. The three-level scale allowed us to develop a finer grained understanding of the quality of the decisions. 

\begin{lstlisting}[captionpos=b,language=Java, caption=A simple decision from Hadoop v. 0.9.0, label=lst:d1,language=java,basicstyle=\footnotesize, 
frame=single, linewidth=0.49\textwidth, xleftmargin=0.025\textwidth, xrightmargin=0.025\textwidth]
Rationales:
 Issue 1:
   Desc: Seperating user logs from system 
                logs in map reduce
   ID  : HADOOP-489
Consequences:
 Change 1: 
   Added: org.apache.hadoop.mapred.TaskLog
\end{lstlisting}


\begin{lstlisting}[captionpos=b, caption=Part of a crosscutting decision from Hadoop v. 0.10.1, label=lst:d2,language=java,basicstyle=\footnotesize, 
frame=single, linewidth=0.49\textwidth, xleftmargin=0.025\textwidth, xrightmargin=0.025\textwidth]
Rationales:
 Issue 1:
   Desc: Implement the LzoCodec to support  
         the lzo compression algorithms
   ID  : HADOOP-851
 Issue 2:
   Desc: Native libraries are not loaded
   ID  : HADOOP-873
Consequences:
 ...
\end{lstlisting}

As illustrative examples, we explain the scoring procedures for two decisions in Hadoop. Listing \mbox{\ref{lst:d1}} displays a simple design decision as uncovered by \mbox{\ourappr} in Hadoop version 0.9.0. The rationale consists of a single issue that explains the intent is to separate the user logs from system logs. However, the rationale summary does not explain why this needs to happen. Looking at the issue in Jira, the reason is that system logs are cluttering the user logs, and system logs need to be cleared out more frequently than user logs. Since we had to look at the issue to understand ``why'' this decision was made, the \mbox{\emph{Rationale Clarity}} in this case was scored 0.5. Since we only have one issue, the \mbox{\emph{Rationale Cohesion}} is not applicable. The consequence involves one change with a single architectural delta, i.e., adding the \texttt{TaskLog}. The relationship of this change to the issue is clear and the change size is tractable. Therefore, \mbox{\emph{Consequence-Rationale Association}} and \mbox{\emph{Consequence Tractability}} each  received 1.
In Listing \mbox{\ref{lst:d2}} which is a crosscutting example from Hadoop 0.10.1, although the rationales seem unrelated, after inspecting the code and issue logs we realized that \mbox{\emph{LzoCodec}} will be available only if the \mbox{\emph{Native Library}} is loaded. Therefore, this decision received 0.5  for \mbox{\emph{Rationale-Cohesion}}.

\begin{figure}[]
	\vspace{-2mm}
	\centering
	\includegraphics[width = .9\columnwidth]{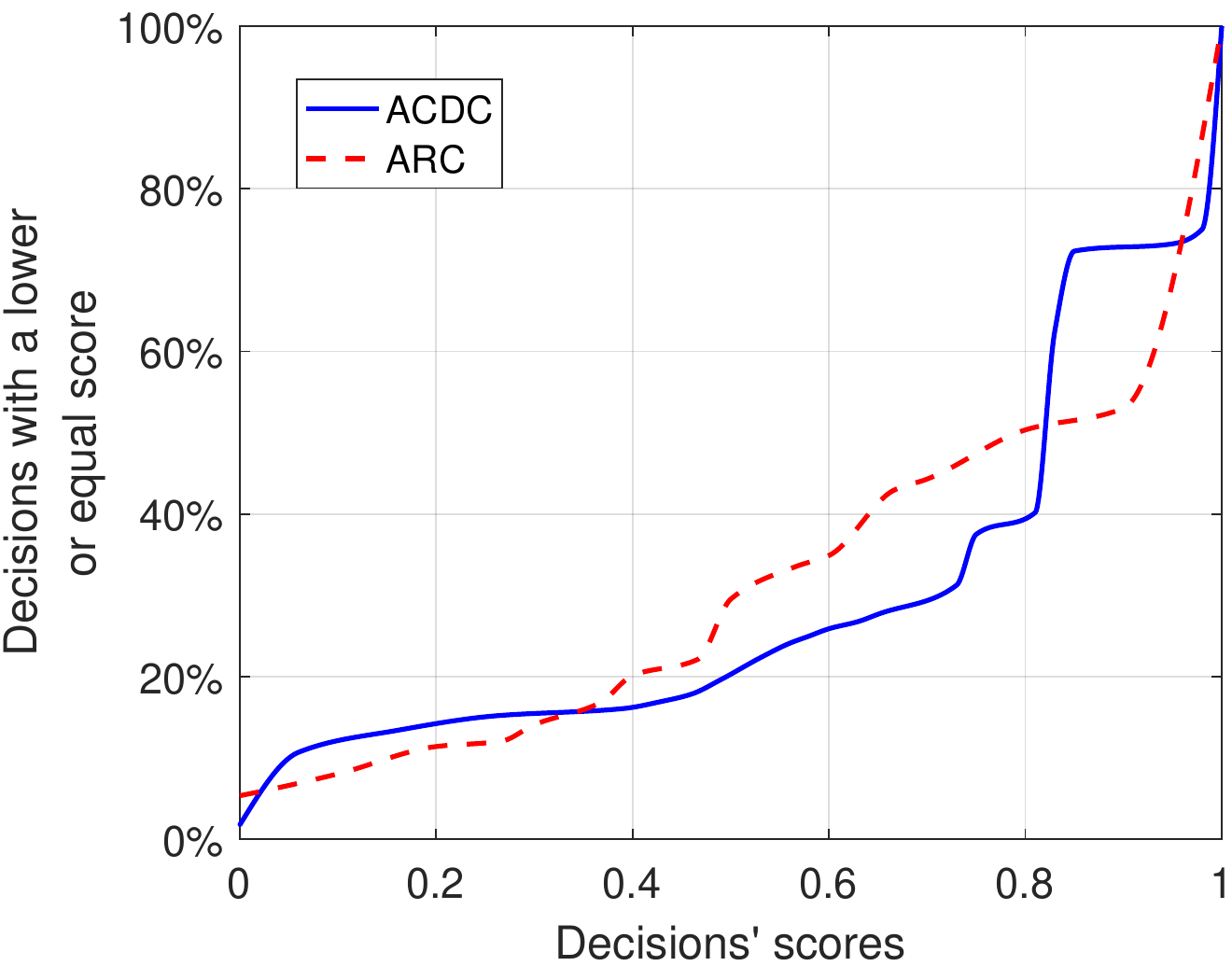}
	\caption[]{Smoothed cumulative distribution of the decision scores for Hadoop. }
	\label{fig:hadoop_ecdf}
\end{figure}
\begin{figure}[]
	\vspace{-2mm}
	\centering
	\includegraphics[width = .9\columnwidth]{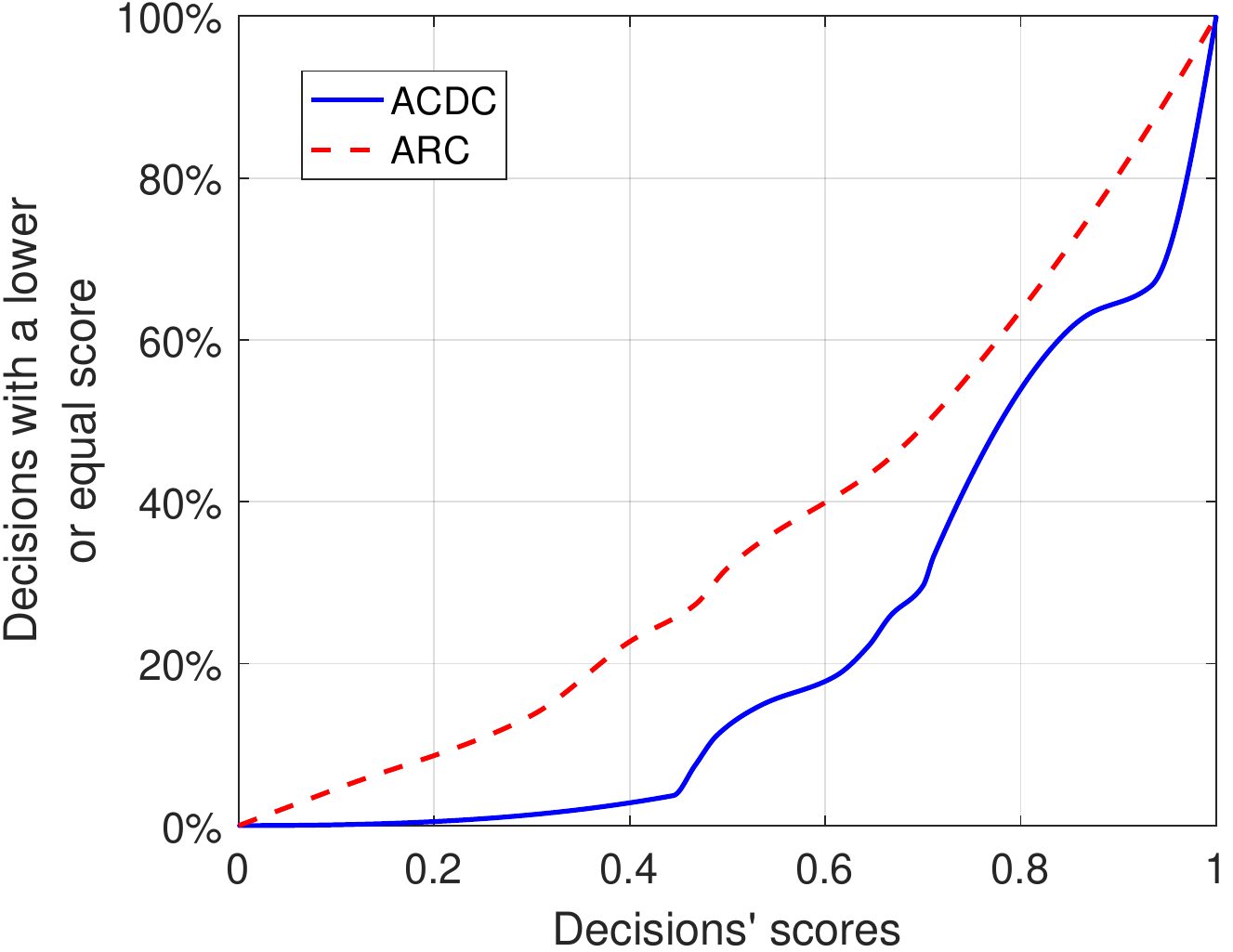}
	\caption[]{Smoothed cumulative distribution of the decision scores for Struts. }
	\label{fig:struts_ecdf}
\end{figure}

Table \ref{tab:averages} displays the average scores of the analyzed decisions, grouped by the decision type and the recovery technique used for uncovering the decisions. 
Figures \ref{fig:hadoop_ecdf} and \ref{fig:struts_ecdf} display the cumulative distributions of the decision scores for Hadoop and Struts, respectively. The right-leaning feature of these distributions indicates that the higher-quality decisions are more prevalent than the lower-quality ones. 
The threshold of acceptability for measuring precision is adjustable, but in our evaluation we required  that a decision scores at least 0.5 in the majority (i.e., at least three) of the criteria. In our analyses, on average (considering both \mbox{\arc} and \mbox{\acdc}) \mbox{76\%} of the decisions for Hadoop and \mbox{78\%} of the decision for Struts met this condition. 


Most of the unacceptable decisions were made in the newly introduced major versions of the two systems. In our previous work, we  observed that the number of architectural changes between a minor version (e.g., 0.20.2) and the immediately following  version that is major (1.0.0) tends to be significantly higher than the architectural change between two consecutive minor versions. In these cases, the decision sizes (number of rationales and consequences) tend to be higher than our conservative thresholds, and these decisions tend to be rated as unacceptable. However, these decisions still provide valuable insight into why the architecture has changed. 

The reason that the \arc-based decisions generally score lower (i.e., they are less right leaning) than the \acdc-based ones  is due to the nature of  changes extracted by \arc. While \acdc adopts primarily a structural approach to architecture, \arc follows a semantic approach and requires a higher level of system understanding. Therefore, attaining a conclusive rating for these decisions was not possible by only looking at the decision elements defined earlier.  
Our findings suggest that the uncovered decisions based on \arc are more suitable for experienced users. 

\subsection{Recall}\label{sec:recall}
Our objective in assessing the recall of our approach is to find out the extent to which \ourappr manages to successfully capture the design decisions in our subject systems. 
Based on the definition of the architectural design decisions (recall Section \ref{sec:foundation}), every architectural change is a consequence of a design decision. Therefore, we use the coverage of architectural changes by the identified design decisions as a proxy indicator for measuring the recall of \ourappr.  

\begin{table}[]
	\centering
	\caption{Architectural Change  Coverage}
	\label{tab:coverages}
	\begin{tabular}{@{}lcccc@{}}
		\toprule
		\multirow{2}{*}{\textbf{}} & \multicolumn{2}{c}{\textbf{Hadoop}} & \multicolumn{2}{c}{\textbf{Struts}} \\ \cmidrule(l){2-5} 
		                                  & \textbf{acdc}    & \textbf{arc}        & \textbf{acdc}  & \textbf{arc} \\ \midrule
		\textbf{Before Cleanup}     & 0.20             &  0.19    & 0.21     &    0.24                    \\
		\textbf{After Cleanup}      & 0.85             &  0.67    & 0.80     &    0.63                    \\
		\bottomrule
	\end{tabular}
\end{table}

Our initial analyses reported an underwhelming recall: only a relatively small fraction of  the extracted changes  formed design decisions. The first row of Table \ref{tab:coverages} displays the results of our initial analyses. 
The recall of the extracted architectural changes was consistently around 20\% in our subject systems regardless of the used recovery technique. 
To understand the root cause of this, we manually examined the detected architectural changes for which \ourappr could not locate the rationale. 
We were able to identify two major reasons why  an architectural change was not marked as part of a design decision by \ourappr. 
The first was when architectural change was happening in the off-the-shelf components that are integrated with the system and evolve separately. These can be third-party libraries, integrations with the other Apache software projects, or even changes in the core Java libraries that are detected by the recovery techniques. Examples of this phenomenon for Struts includes changes to the Spring Framework's architecture \cite{johnson2004spring}, and for Hadoop  changes to Jetty \cite{jetty} and several non-core Apache Common projects. The second reason is what we call the ``orphaned commit'' phenomenon. Orphaned commits are the commits that conceptually belong to an issue, but (1) were not added to an issue, (2) have been merged with the code-base before their containing issues has been marked as resolved, or (3) a human error in the issue data rendered them useless for our approach (e.g., incorrectly specified affected version).

We consider orphaned commits a shortcoming of our approach that can affect the recall. However, the imposed changes on a system's architecture do not capture the original intentions of the developers and architects. 
Therefore, we carefully inspected the architectural changes to eliminate the ones caused by external factors. 
In our inspection, we created a list of namespaces whose elements should not be considered  architectural changes caused by the developer decisions. Truncated lists of these namespaces for Hadoop and Struts are displayed in Listings \ref{lst:hadoop} and \ref{lst:struts}, respectively. We verified each entry by searching the system's code repository and confirming that the instances where imported and not developed internally by the developer teams.

\begin{lstlisting}[captionpos=b,language=Java, caption=Imported namespaces for Hadoop, label=lst:hadoop,language=java,basicstyle=\footnotesize, 
frame=single, linewidth=0.49\textwidth, xleftmargin=0.025\textwidth, xrightmargin=0.025\textwidth]
com.facebook.*
java.lang.*
org.apache.commons.cli.*
javax.ws.rs.*
...
\end{lstlisting}
\begin{lstlisting}[captionpos=b,language=Java, caption=Imported namespaces for Struts, label=lst:struts,language=java,basicstyle=\footnotesize, 
frame=single, linewidth=0.49\textwidth, xleftmargin=0.025\textwidth, xrightmargin=0.025\textwidth]
com.opensymphony.xwork2.util.*
java.io.*
org.apache.commons.*
org.springframework.*
...
\end{lstlisting}


We then reevaluated the recall of the changes for our approach. The results are displayed in the second row of  Table \ref{tab:coverages}.
The recall of our approach after eliminating externally caused changes is \textbf{73\%} on average. This also reveals an interesting byproduct of \ourappr, namely, by using \ourappr or a specially modified version of it, we can detect the parts of a system that are not developed or maintained by the system's core team. This information can be used for automatic detection of external libraries and dependencies in software systems, and can help the recovery techniques in extracting a more accurate view of a system's ``core'' architecture. 




\section{Threats to validity}\label{sec:threats}
We identify several potential threats to the validity of our
approach and results with their corresponding mitigating factors.
The key threats to \textbf{external validity} involve our subject
systems. Although we use two systems for our evaluations, 
these systems were chosen from the higher end of the Apache spectrum in terms of size and lifespan, each have a vibrant community, and are widely adopted. 
Another threat stems from the fact that both of our systems are using GitHub and Jira. However, \ourappr only relies on the basic issue and commit information that can be found in any generic issue tracker or version control system.
The different numbers of versions analyzed per
system pose another potential threat to validity. This is
unavoidable, however, since some systems simply undergo
more evolution than others.

The \textbf{construct validity} of our study is mainly threatened by
the accuracy of the recovered architectural views and of our
detection of architectural decisions. To mitigate the first threat,
we selected the two architecture recovery techniques, ACDC
and ARC, that have demonstrated the greatest accuracy in our
extensive comparative analysis of available techniques \cite{garcia2013comparative}.
These techniques are developed independently of one another and use very different
strategies for recovering an architecture. This, coupled with the
fact that their results exhibit similar trends, helps to strengthen
the confidence in our conclusions. 
The manual inspection of the accuracy of the design decisions uncovered by our approach is another threat. Human error in this process could affect the reported results. To alleviate this problem, two PhD students independently analyzed the results to limit the potential biases and mistakes. Moreover, the inspection procedure was designed to be very conservative.
\section{Related Work}\label{sec:related}

We will briefly touch upon the most closely related approaches that have been proposed to justify, model, or recover architectural design decisions.

Jansen and Bosch et al. \cite{jansen2005software, bosch2004software} defined architectural design decisions and argued for the benefits of the invaluable information getting lost when architecture are modeled using purely structural elements. 
Kruchten et al. proposed an ontology that classified architectural decisions into 3 categories: (1) existence decisions (ontocrises), (2) property decisions (diacrises), and (3) executive decisions (pericrises) \cite{kruchten2004ontology}. 
Taylor et al. \cite{taylor2009software} also defined software architectures in terms of the ``principal design decisions'' yielding them.
Several researchers focused on studying the concrete benefits of using design decisions in improving software system's quality \cite{tang2008design}, and decision making under uncertainty \cite{burge2008design}. Falessi et al. extensively studied design rationale and argued for the value of capturing and explicitly documenting this information \cite{falessi2013value}. A recent expert survey by Weinreich et al. \cite{weinreich2015expert} showed that knowledge vaporization is a problem in practice, even at the individual level. 



Roeller et al. \cite{roeller2006recovering} proposed RAAM to support reconstruction of the assumptions picture
of a system. Assumptions are the architectural design decisions that are made during the evolution of a software system. 
A serious problem with their approach is that the researchers need to acquire a deep understanding of the software system
ADDRA \cite{jansen2008documenting} was designed to recover architectural
design decisions in an after the fact documentation effort. It was built on the premise that in practice, software architectures are often documented after the fact, i.e. when a system is realized and architectural design decisions have been taken. Similar to RAAM, ADDRA also relies on the tacit knowledge of the architect. 


\section{Conclusions and Future Work}\label{sec:conclusions}
Modeling software systems based on the original four ``C''s---components, connectors, configurations, and constraints---that are considered the structural building blocks of a system's architecture 
is deemed responsible for a phenomenon called knowledge vaporization. Knowledge vaporization in software systems plays a major role in increasing maintenance costs, and also exacerbates architectural drift and erosion \cite{jansen2005software}. To that end, researchers have more recently tried to approach architectures from the perspective of architectural design decisions. These efforts, however, have lacked a precise understanding of architectural design decision's \emph{definition}, \emph{identification}, \emph{classification}, \emph{reification}, and \emph{evolution}.

To address these questions, we have developed \ourappr, a technique for automatically uncovering architectural design decisions in existing software systems. We built our definition of an architectural design decision on the constructs previously identified by the research community. 
Our efforts have led to the first automated decision recovery technique that relies solely on the information in issue and code repositories of a software system which for many software systems are the only reliable source of information. 
Our empirical evaluation shows that \ourappr exhibits high accuracy and recall and can successfully detect different types of design decisions.

There are a number of remaining research challenges that will guide our future work. There is a slew of information in software repositories that can help increase the accuracy of our approach. These include comments, commit messages, documentations, pull requests, etc. 
Our approach can be used in tandem with other techniques aiming to support better understanding of quality repercussion of architectural changes \cite{wicsa,shahbazian2016end,Safi:2015:DEA:2786805.2786836}. \ourappr can also be extended with a neural abstractive summarization technique \cite{rush2015neural} to provide more accurate summaries of the rationales and consequences.

\balance

\bibliographystyle{abbrv}
\bibliography{icsa-decision}

\end{document}